\documentclass[superscriptaddress,twocolumn,showpacs,preprintnumbers,amsmath,amssymb,prl]{revtex4}

\usepackage{graphicx}
\usepackage{dcolumn}
\usepackage{bm}

\begin{document}

%\preprint{APS/123-QED}

\title{Spin dependent impurity effects in the 2D frustrated magnetism of NiGa$_2$S$_4$}

\author{Yusuke Nambu}
\affiliation{Department of Physics, Kyoto University, Kyoto 606-8502, Japan}
\affiliation{Institute for Solid State Physics, University of Tokyo, Kashiwa, Chiba 277-8581, Japan}
\author{Satoru Nakatsuji}
\affiliation{Institute for Solid State Physics, University of Tokyo, Kashiwa, Chiba 277-8581, Japan}
\author{Yoshiteru Maeno}
\affiliation{Department of Physics, Kyoto University, Kyoto 606-8502, Japan}
\author{Edem K. Okudzeto}
\affiliation{Department of Chemistry, Louisiana State University, Baton Rouge, LA 70803}
\author{Julia Y. Chan}
\affiliation{Department of Chemistry, Louisiana State University, Baton Rouge, LA 70803}

\date{\today}

\begin{abstract}
Impurity effects on the triangular antiferromagnets Ni$_{1-x}M_x$Ga$_2$S$_4$ ($M=$ Mn, Fe, Co and Zn) are studied.
The 2D frozen spin-disordered state of NiGa$_2$S$_4$ is stable against the substitution of Zn$^{2+}$ ($S=0$) and Heisenberg Fe$^{2+}$ ($S=2$) spins, and exhibits a $T^2$-dependent magnetic specific heat, scaled by the Weiss temperature.
In contrast, the substitutions with Co$^{2+}$ ($S=3/2$) spin with Ising-like anisotropy and Heisenberg Mn$^{2+}$ ($S=5/2$) spin induce a conventional spin glass phase.
From these comparisons, it is suggested that integer size of Heisenberg spins is important to stabilize the 2D coherent behavior observed in the frozen spin-disordered state.
\end{abstract}

\pacs{75.40.Cx, 75.50.Ee}

\maketitle

Geometrically frustrated and low dimensional magnets have attracted considerable interest because of the possible emergence of the novel collective phenomena upon suppression of conventional magnetic orders.
One of the most prominent examples of such phenomena is the gap formation in 1D antiferromagnetic (AF) chains \cite{Haldane}.
The so-called Haldene gap appears only for chains with integer spins.
In higher dimensions, the possibility of such a 2$S$ parity dependent ground state is interesting and indeed have been conjectured by theorists \cite{sse}, but no experimental realizations of such an effect has been reported.

In 2D antiferromagnets (AFMs) based on arrays of triangles, geometrical frustration in addition to reduced dimensionality may favor collective behavior without spin order.
One of such exotic phenomena is the quadratic temperature dependence in the specific heat without long range order, anticipated in a system with 2D coherent spin excitations.
This phenomenon has been observed in an increasing number of AFMs such as the Kagom\'e AFMs, SrCr$_{9p}$Ga$_{12-9p}$O$_{19}$ \cite{SCGO}, deuteronium jarosite \cite{jarosite} and more recently in the triangular AFM NiGa$_2$S$_4$ \cite{Nigas}.
However, no clear explanation has been made for the mechanism to stabilize such a coherent behavior in the absence of long range order.

Here we report the spin dependent impurity effects in the 2D coherent behavior in NiGa$_2$S$_4$, the first example of the spin $S=1$ 2D AFMs with an exact triangular lattice \cite{Nigas}.
Our results strongly suggest the 2$S$ parity dependence in the 2D frustrated magnetism.
Interestingly, this magnet does not form a conventional 3D AF order at least down to 0.08 K in spite of the fact that its AF interactions have the energy scale of about 80 K.
Instead, $^{69,71}$Ga nuclear quadrupole resonance and muon experiments have clarified unusual bulk spin freezing across $T^{\ast} = 10$ K that has a highly extended critical regime down to 2 K \cite{NQR}.
Below 2 K, neutron measurements have revealed quasi-2D noncollinear correlation, whose spin-spin inplane correlation length stays finite around 7 times the lattice spacing, while the interplane correlation is so weak that it barely reaches the nearest neighbor planes.
Moreover, inhomogeneous internal field at the Ga site was observed, and the nuclear lattice relaxation rate shows nearly cubic temperature dependence below 1 K that suggests a 2D magnon-like dispersive mode.
Further consistent result has been obtained for the magnetic specific heat ($C_{\rm M}$) that shows the quadratic temperature dependence below about 4 K, suggesting an AF spin-wave-like mode in 2D despite the absence of a long range AF order.

Thus, a major question is the origin of the 2D spin-wave-like coherent behavior. 
In order to deepen our understanding, we have studied magnetic and nonmagnetic impurity effects on the thermodynamic properties of NiGa$_2$S$_4$.
Our experiments using both single crystals and polycrystalline samples of the insulating Ni$_{1-x}M_x$Ga$_2$S$_4$ ($M=$ Mn, Fe, Co and Zn) indicate that the Fe$^{2+}$ ($S=2$) and Zn$^{2+}$ ($S=0$) substituted systems, which have Heisenberg and integer spins, surprisingly retain the $T^2$-dependent $C_{\rm M}$ that is scaled with $|\theta_{\rm W}|$.
In contrast, this is not the case for the Co and Mn substitutions.
The conventional spin glass phase emerges with the substitution of impurities with half-odd integer spins, Ising-like Co$^{2+}$ ($S=3/2$) and Heisenberg Mn$^{2+}$ ($S=5/2$) spins.
This suggests that integer size of Heisenberg spins is important to stabilize the 2D spin-wave-like coherent behavior observed in the unusual frozen spin disordered state.

The polycrystalline samples and single crystals of Ni$_{1-x}M_{x}$Ga$_2$S$_4$ are synthesized by solid state reaction and chemical vapor transport methods, respectively \cite{ICCG}.
In order to obtain a homogeneous mixture of Ni and $M$, we first ground the same molar amount of Ni and $M$ powder, and repeated the dilutions by adding equal molar amount of Ni powder each time until obtaining the appropriate $M$ concentration.
The amount of $M$ in single crystals was determined by energy-dispersive X-ray analyses.
Powder X-ray diffraction at room temperature on polycrystalline samples indicates single phase isostructural to NiGa$_2$S$_4$ at least up to $x=0.1$ for Mn, $x=1$ for Fe, $x=0.5$ for Co and $x=0.3$ for Zn.
Previous studies have clarified that both Zn and Fe atoms occupy the Ni site \cite{Zn,Fegas}.
For $M$ = Co and Mn, single crystal X-ray diffraction experiments were performed, and confirmed that Co atoms are indeed located at the Ni site.
For Mn, the substitution cannot be directly determined because of the low concentration.
However, the refined cell obtained for Mn is isostructural with the trigonal NiGa$_2$S$_4$ and consistent with lattice parameters obtained from powder diffraction.
With Pauling's electrostatic valency principle \cite{Pauling}, one expects Mn atoms to occupy Ni rather than Ga site because of the difference in ionic radii (0.66 \AA\ and 0.47 \AA (Coordination Number, CN = 4) for Mn$^{2+}$ and Ga$^{3+}$, respectively).
Furthermore, the oxidation state as obtained from X-ray photoemission spectroscopy (XPS) and susceptibility measurements is consistent with the Mn$^{2+}$ state.
DC magnetization $M$ was measured using a SQUID magnetometer.
The specific heat $C_P$ was measured by thermal relaxation method.
The lattice contribution $C_{\rm L}$ is estimated using the same method as in Ref. \cite{Zn}. 
Among the impurity ions, Mn and Co have nuclear spins of 5/2 and 7/2, respectively.
The Co-nuclear quadrupole contribution, $C_{\rm nq}=6.11\times 10^{-4}/T^2$ (J/mol K), was estimated by a standard analysis of the field dependence.
The same procedure was not applicable for the Mn case because of the lack of the data points at low enough temperatures.
Instead, we estimated the nuclear quadrupole contribution $C_{\rm nq}=8.55\times 10^{-4}/T^2$ (J/mol K) by assuming that Mn is located under the same crystal field as Co.
Finally, $C_{\rm M}$ was estimated using the relation, $C_P=C_{\rm L}+C_{\rm nq}+C_{\rm M}$.

The high spin (HS) state of Ni$^{2+}$ in pure NiGa$_2$S$_4$ and Mn$^{2+}$ has the configuration of $(t_{2g})^6(e_g)^2$ and  $(t_{2g})^3(e_g)^2$, respectively.
Both have no orbital degree of freedom, and thus their spins should be of Heisenberg type.
Fe$^{2+}$ has the $(t_{2g})^4(e_g)^2$ configuration with HS $S=2$, and thus is Jahn-Teller (JT) active.
A weak rhombohedral distortion should split the $t_{2g}$ orbitals and stabilize one $a_{1g}$ level against the two $e_g^{\prime}$ levels by a JT gap ($\Delta_{\rm JT}$), which should be of the order of 300 K \cite{Fegas}.
Thus, the ground state has no orbital degree of freedom, and the spins should be of Heisenberg type.
For Co$^{2+}$ ($(t_{2g})^5(e_g)^2$), there is residual degeneracy of the $e_g^{\prime}$ orbitals, which should lead to Ising-like anisotropy of spins, as observed in CoCl$_2\cdot$2H$_2$O \cite{CoIsing}.

Figures 1(a) and (b) show the temperature dependence of the susceptibility $\chi\equiv M/H$ and $C_P/T$ for various samples.
\begin{figure}[t]
\includegraphics[width=250pt]{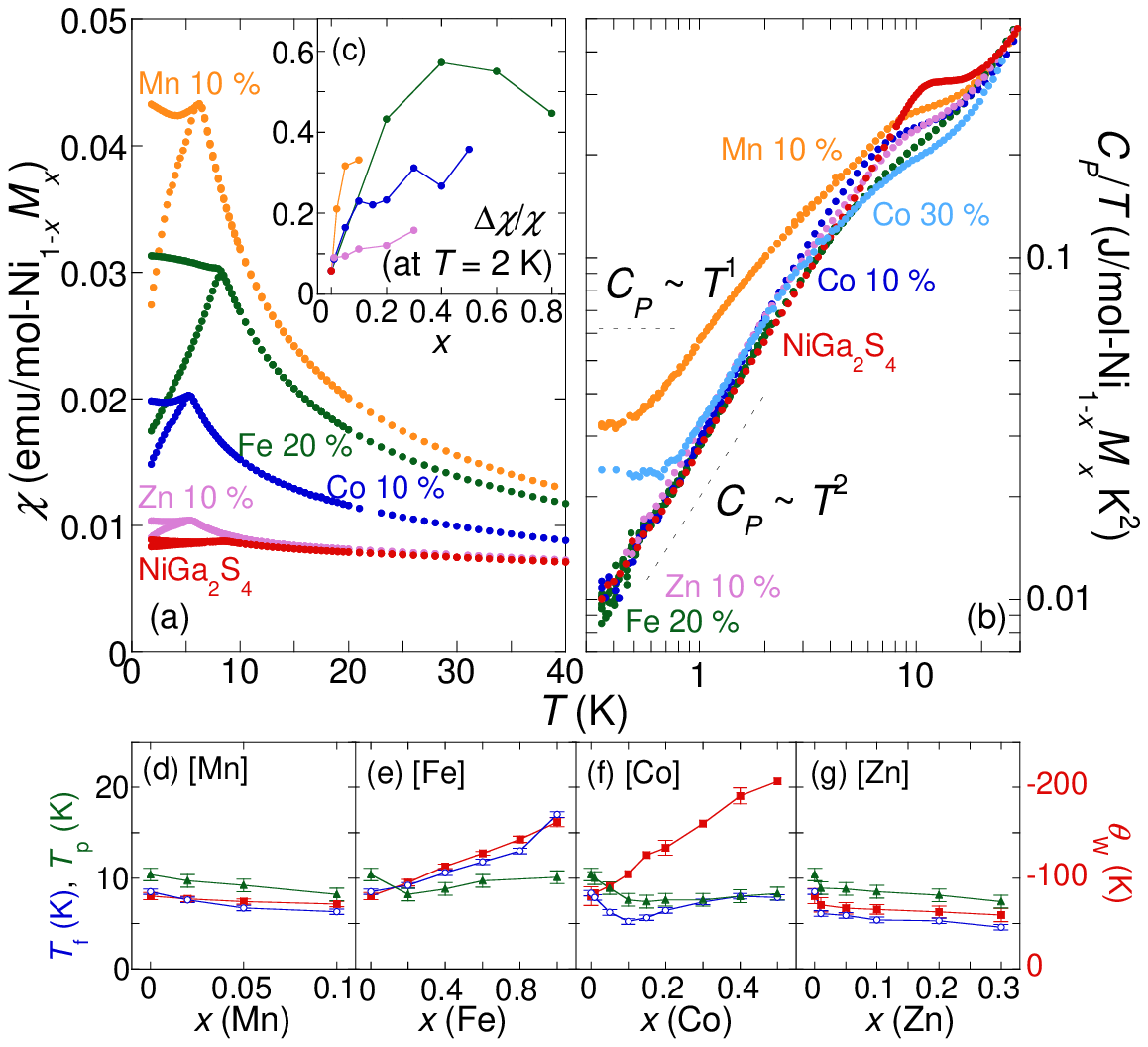}
\caption{(color online) Temperature dependence of (a) the susceptibility under $\mu_0 H=0.01$ T, and (b) $C_P/T$ under zero field observed by using polycrystalline samples. Impurity concentration dependence of (c) $\Delta\chi/\chi$ at $T=2$ K and (d,e,f,g) $T_{\rm f}$ (blue solid square), $T_{\rm p}$ (green solid triangle) and $\theta_{\rm W}$ (red open circle) for all the substituted systems.}
\end{figure}
All compounds exhibit spin freezing under $\mu_0 H=0.01$ T as evident from the bifurcation between field-cooled (FC) and zero field-cooled (ZFC) sequences.
Pure NiGa$_2$S$_4$ also exhibits a small bifurcation ($\sim 6$ \%) below $T_{\rm f}=8.5$ K.
In Fig.1(b), despite the spin freezing, Fe 20 \% and Zn 10 \% substituted compounds surprisingly retain the same $T^2$-dependence of $C_{\rm M}$ as NiGa$_2$S$_4$, while $C_{\rm M}/T$ becomes almost constant at low temperatures for Mn 10 \% and Co 30 \% substitution.

The inset, Fig.1(c), presents impurity concentration ($x$) dependence of the ratio ($\Delta\chi/\chi$) between the size of the bifurcation ($\Delta\chi$) and the FC susceptibility ($\chi$) at $T=2$ K.
All the impurities enhance $\Delta\chi/\chi$ with substitution.
Especially for the Fe substitution, it has a maximum value around $x = 0.5$, suggesting that $\Delta\chi$ is enhanced by randomness caused by substitution.
Thus, the bifurcation observed in NiGa$_2$S$_4$ below $T_{\rm f}=8.5$ K may be attributed to freezing of minority spins following the bulk spin freezing at $T^{\ast}=10$ K.
While the increase in $\Delta\chi$ with substitution is similar to what has been observed in conventional spin glass systems, a clear difference appears in the change in $T_{\rm f}$.
Figures 1(d), (e), (f) and (g) show $x$-dependence of $T_{\rm f}$, peak temperature $T_{\rm p}$ of $C_{\rm M}/T$, and $\theta_{\rm W}$, where $T_{\rm f}$ is determined to be the bifurcation point under 0.01 T and $\theta_{\rm W}$ is by the Curie-Weiss (CW) analysis using the formula $\chi=C/(T-\theta_{\rm W})$.
Normally, for conventional spin glasses, $T_{\rm f}$ increases with the impurity concentration.
However, $T_{\rm f}$ for all the impurity substitutions except Fe decreases in the small concentration regime ($x\le 0.1$).
Moreover, for the Fe substituted system, $T_{\rm f}$ decreases with Ni substitution for FeGa$_2$S$_4$, and this is not consistent with conventional spin glass behavior.
Instead, the Fe, Zn and Mn substituted systems show a scaling behavior, $|\theta_{\rm W}|\sim 10T_{\rm f}$, indicating that $T_{\rm f}$ is determined by $\theta_{\rm W}$.
However, this is not the case only for the Co substituted systems.
Meanwhile, $T_{\rm p}$ roughly scales with $T_{\rm f}$ but always slightly above $T_{\rm f}$, suggesting the broad peak in $C_{\rm M}/T$ is associated with the spin freezing.
Only one exception for this case is the Fe substituted system.
This is because of the existence of $\Delta_{\rm JT}$, which will be discussed later.

Effective moments obtained by the CW fitting to the susceptibility data of polycrystalline samples are plotted in Fig.2(a) and (b).
\begin{figure}[t]
\includegraphics[width=250pt]{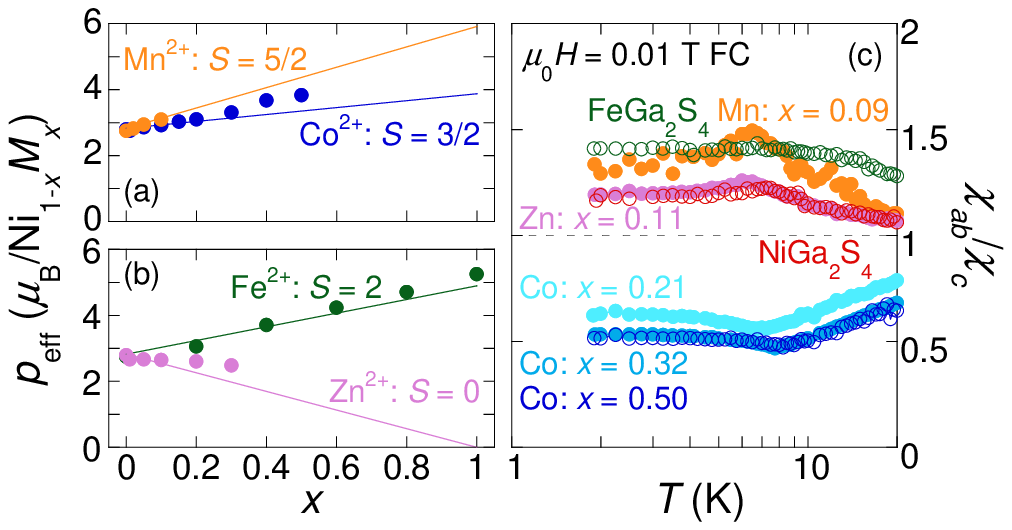}
\caption{(color online) (a,b) Effective moments and spin size of $M$ in the polycrystalline samples. (c) Temperature dependence of the ratio between $\chi_{ab}$ and $\chi_c$.}
\end{figure}
The results are consistent with the value (solid line), $2\sqrt{2}(1-x)+2\sqrt{S_{\rm imp}(S_{\rm imp}+1)}x$, estimated for HS state, where $S_{\rm imp}$ is the size of impurity spin.
Furthermore, a systematic change of the lattice constants with substitution \cite{ICCG} and recent XPS measurements by using single crystals \cite{Takubo} also suggest that all the $M$ ions have 2+ valence and are in the HS states.
In Fig.2(a), an additional enhancement of $p_{\rm eff}$ was found in the Co substituted system, which should come from the orbital contribution due to the degeneracy of $e_g^{\prime}$ orbitals.

Figure 2(c) shows anisotropy determined by the ratio between the inplane and interplane susceptibility ($\chi_{ab}$ and $\chi_c$) under 0.01 T FC sequence.
NiGa$_2$S$_4$, FeGa$_2$S$_4$, and the Mn, Zn substituted systems have $1<\chi_{ab}/\chi_c<1.4$, which indicates that spins are of Heisenberg type with weak $XY$ anisotropy, while the Co substituted system shows Ising-like anisotropy with $\chi_{ab}/\chi_c$ down to 0.5.
The observed anisotropy is consistent with the single-ion anisotropy expected for each impurity ion.

For states that have a single energy scale of $|\theta_{\rm W}|$, the molar entropy ($S_{\rm M}$) should take the form $S_{\rm M}=R\ln(2S+1)f(T/|\theta_{\rm W}|)$ including its spin size ($S$).
By taking the derivative of this, one obtains the following relation,
\begin{align}
\frac{{\rm d}S_{\rm M}}{{\rm d}T}=\frac{C_{\rm M}}{T}=\frac{R\ln(2S+1)}{|\theta_{\rm W}|}f^{\prime}(T/|\theta_{\rm W}|).
\end{align}
In order to check the normalization with $|\theta_{\rm W}|$, $C_{\rm M}|\theta_{\rm W}|/(TR\ln(2S+1))$ vs. $T/|\theta_{\rm W}|$ is plotted in Fig.3, where the total molar spin entropy in the vertical axis is given by $R\{(1-x)\ln3+x\ln(2S_{\rm imp}+1)\}$.
\begin{figure}[t]
\includegraphics[width=250pt]{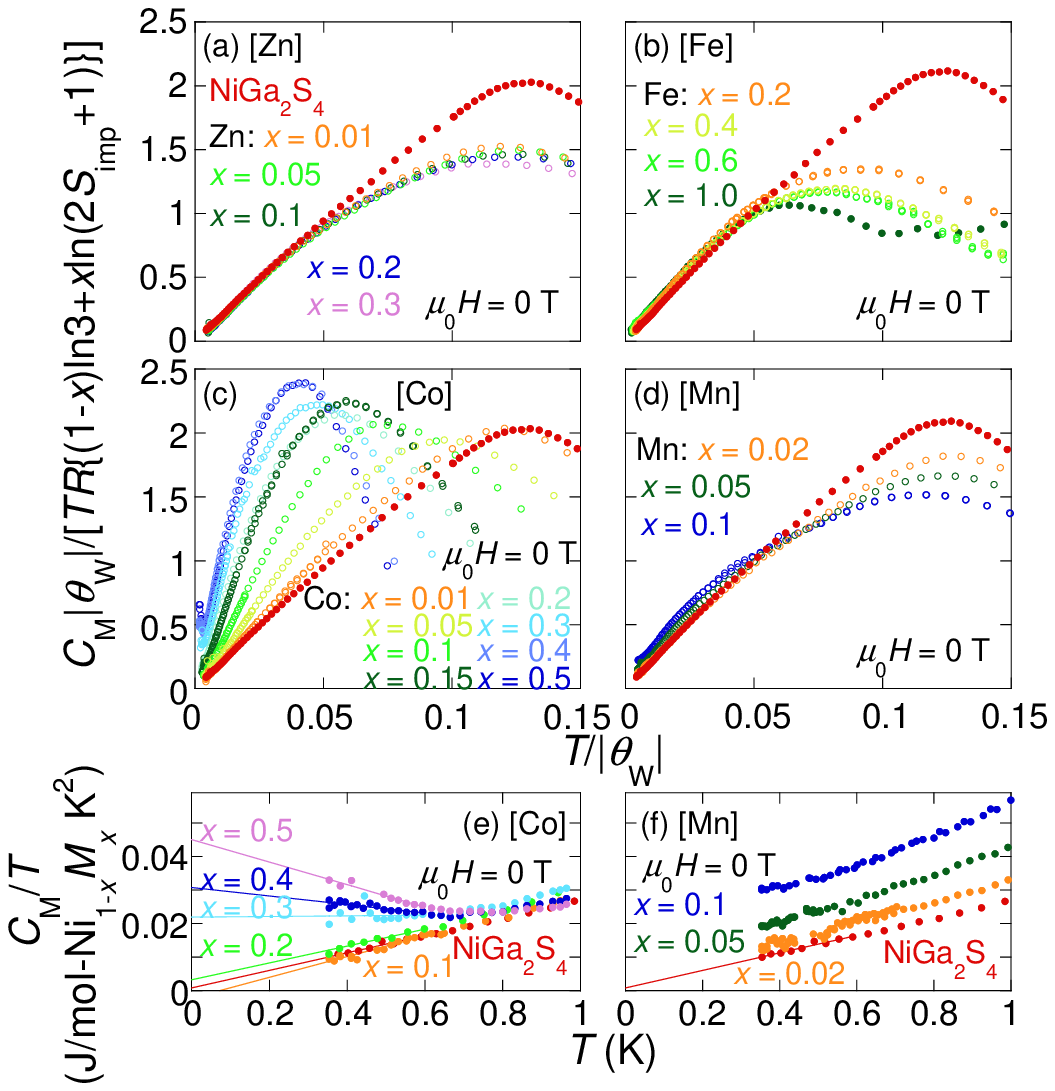}
\caption{(color online) $T/|\theta_{\rm W}|$ dependence of the normalized $C_{\rm M}$ for (a) Zn, (b) Fe, (c) Co and (d) Mn substitutions. $T$-dependence of $C_{\rm M}/T$ for (e) Co and (f) Mn substitutions. Solid lines indicate its linear extrapolation below 0.6 K.}
\end{figure}
Figures 3(a) and (b) are for the Zn and Fe substituted systems.
The Zn substituted system exhibits the $T^2$-dependent behavior of $C_{\rm M}$ up to $x=0.3$.
The substitution leads to the abrupt depression of the peak in $C_{\rm M}/T$ by the substitution of less than 1 \%, indicating our obtained sample of NiGa$_2$S$_4$ is near the clean limit.
The initial slope remain constant, indicating that there exists a normalized function $f(T/|\theta_{\rm W}|)$ and the $T^2$-coefficient of $C_{\rm M}$ is proportional to $|\theta_{\rm W}|^{-2}$.
This $(T/|\theta_{\rm W}|)^2$ form of the specific heat suggests an existence of a 2D spin-wave-like mode whose spin stiffness is proportional to $\theta_{\rm W}$.
The two dimensionality is also consistent with the fact that $\theta_{\rm W}$ should be determined not by the negligibly small interplane coupling, but by the inplane exchange interactions.
For the Fe substitution, all the curves have the well-scaled initial slope up to $T/|\theta_{\rm W}| \sim 0.05$, above which, however, they start deviating from each other.
This may be due to the effect of the JT gap $\Delta_{\rm JT}$ of Fe$^{2+}$ ions.
Given $\Delta_{\rm JT}\sim$ 300 K \cite{Fegas}, the same order of magnitude as $|\theta_{\rm W}|$, the Schottky tail due to $\Delta_{\rm JT}$ might affect the $T$-dependence of $C_{\rm M}/T$ and its peak formation at $\sim T_{\rm p}$.

In Fig.3(c), the normalized $C_{\rm M}$ for the Co substitution is shown.
It is clear that no curves overlap each other, indicating that Co spins with Ising anisotropy completely suppress the scaling behavior of $C_{\rm M}/T$.
Moreover, $\gamma$ which is estimated by the linear extrapolation of $C_{\rm M}/T$ curve below 0.6 K (Fig.3(e)), starts appearing at $x>0.1$ and increases with the Co content.
The anomaly at $x=0.1$ is also seen in $T_{\rm f}$, $T_{\rm p}$ (Fig.1(f)) and $\Delta\chi/\chi$ at 2 K (Fig.1(c)), pointing toward the existence of a crossover at $x\sim 0.1$.
At $x<0.1$, the system exhibits the behaviors not consistent with the conventional spin glass, i.e. $T_{\rm f}$ decreases against impurities, and $\gamma$ keeps its value $\sim 0$ with $C_{\rm M} \sim T^2$-behavior.
At $x>0.1$, on the other hand, the system shows conventional spin glass behavior in the sense that it shows finite $\gamma$ and that both $\gamma$ and $T_{\rm f}$ increase with the substitution.
These results indicate that the ground state of NiGa$_2$S$_4$ is different from that of conventional spin glass state.

Finally, for Mn substitution, even with the relatively low amount ($x\le 0.1$) of impurity, no scaling is found in the normalized $C_{\rm M}/T$ (Fig.3(d)), but a systematic change from the $T^2$dependence of $C_{\rm M}$ (Fig.3(f)).
In order to show this explicitly, we plot in Fig.4 the deviation in $C_{\rm M}$ from the $T^2$-dependence estimated using the scaling law of Eq.(1) as follows.
\begin{figure}[t]
\includegraphics[width=250pt]{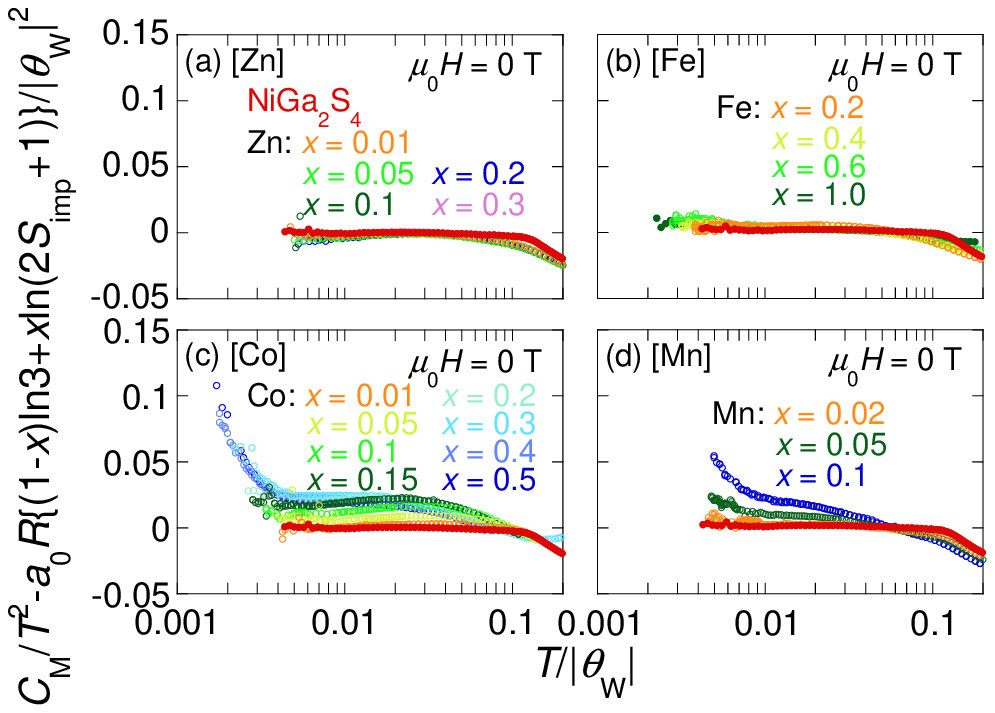}
\caption{(color online) $T/|\theta_{\rm W}|$ dependence of $C_{\rm M}/T^2-A$ for the (a) Zn, (b) Fe, (c) Co and (d) Mn substituted systems. $A$ is determined by using Eq.(2).}
\end{figure}
Suppose $C_{\rm M}=AT^2$ behavior is well scaled by a single energy scale of $|\theta_{\rm W}|$, $A$ for Ni$_{1-x}M_x$Ga$_2$S$_4$ should take the form, 
\begin{align}
A=a_0\frac{R\left\{(1-x)\ln3+x\ln(2S_{\rm imp}+1)\right\}}{|\theta_{\rm W}|^2},
\end{align}
leading $f^{\prime}(T/|\theta_{\rm W}|)=a_0 T/|\theta_{\rm W}|$ in Eq.(1).
Here, the coefficient $a_0=19.4$ is uniquely determined by the $T^2$-coefficient of $C_{\rm M}$ for pure NiGa$_2$S$_4$.
$|\theta_{\rm W}|$ is determined by the CW analysis of $\chi(T)$, thus there is no free parameter in this analysis.
The vertical axis in Fig.4 presents $C_{\rm M}/T^2-A$, using $A$ estimated by Eq.(2).
If the scaling works, the vertical axis in Fig.4 should take constant zero value.
Indeed, this is the case for the Zn and Fe substitution (Figs. 4(a) and (b)).
However, the Mn and Co substituted systems do not hold the $C_{\rm M}\sim T^2$ behavior (Figs. 4(c) and (d)), but have a finite $\gamma$.
These indicate that a conventional spin glass phase is stabilized by the Mn substitution, as for the Co case with $x({\rm Co})>0.1$.

To summarize the experimental results, a conventional spin glass phase emerges with the substitution of magnetic impurities with half-odd integer spins: Ising-like Co spins and Heisenberg Mn spins.
Such conventional spin glass behavior is expected, given that the randomness due to impurities and geometrical frustration inherent in the triangular lattice.
Surprisingly, however, the Fe and Zn substituted systems which have Heisenberg and integer spins, retain the $T^2$-dependent $C_{\rm M}$, that is scaled with $|\theta_{\rm W}|$.
Moreover, $T_{\rm f}$ also scales with $\theta_{\rm W}$.
These suggests that the unusual spin frozen state of NiGa$_2$S$_4$ is not conventional spin glass, and its 2D coherent behavior represented by $C_{\rm M}\sim (T/|\theta_{\rm W}|)^2$ appears only for integer Heisenberg spins.

Two types of theoretical proposals have been made for the origin of the 2D coherent behavior in NiGa$_2$S$_4$. First one is the spin nematic state that describes unconventional long-range order by magnetic quadrupoles without long-range two-spin correlation.
In this state, the site average of spin is zero \cite{Tsunetsugu,Penc,Senthil,Shen}, which is inconsistent with observed results such as the internal field.
Second one is a Kosterlitz-Thouless (KT) type phase driven by $Z_2$ vortices \cite{KY}.
In this case, the anomaly in $C_{\rm M}$ is due to the condensation of $Z_2$ vortices that are formed by vector spin chirality.
It is likely that Ising type impurity spins destabilize such crossover as in our experiment because vector spin chirality can be only defined for the Heisenberg spins.
A recent theory \cite{Fujimoto} based on quantum calculations that treats the low-$T$ region below the KT type crossover predicts the $(T/|\theta_{\rm W}|)^2$-dependent specific heat and the constant susceptibility as $T\rightarrow 0$ K, which is consistent with our observations. 

However, this $Z_2$ vortex mechanism alone cannot explain the 2$S$ parity dependence observed in the present experiments.
The distinct behavior between integer and half-integer spins may be understood as a quantum effect, for example, through quadratic form of spin exchange interactions, as known in Haldane systems in 1D \cite{AKLT}.
Indeed such a quadratic form of spin interactions has been discussed to explain the magnetism of the related compound NiS$_2$ \cite{MSE}.
This type of quantum effect might be also the origin of the finite inplane correlation length.

We acknowledge K. Ishida, K. Onuma, H. Kawamura, K. Takubo, T. Mizokawa, H. Tsunetsugu, S. Fujimoto and A. Tanaka for valuable discussions.
This work was supported in part by Grant-in-Aid for Scientific Research on Priority Areas (190520019052003).


\begin{thebibliography}{99}
\bibitem{Haldane} I. Affleck, J. Phys. Condens. Matter {\bf 1}, 3047 (1989).
\bibitem{sse} F.D.M. Haldane, Phys. Rev. Lett. {\bf 61}, 1029 (1988); A. Tanaka and X. Hu, Phys. Rev. Lett. {\bf 95}, 036402 (2005).
\bibitem{SCGO} A.P. Ramirez, G.P. Espinosa and A.S. Cooper, Phys. Rev. Lett. {\bf 64}, 2070 (1990). 
\bibitem{jarosite} A.S. Wills {\it et al}, Europhys. Lett. {\bf 42}, 325 (1998).
\bibitem{Nigas} S. Nakatsuji {\it et al}., Science {\bf 309}, 1697 (2005).
\bibitem{NQR} H. Takeya {\it et al}., Phys. Rev. B {\bf 77}, 054429 (2008).
\bibitem{ICCG} Y. Nambu {\it et al}., J. Cryst. Growth (to be published).
\bibitem{Zn} Y. Nambu, S. Nakatsuji, and Y. Maeno, J. Phys. Soc. Jpn. {\bf 75}, 043711 (2006).
\bibitem{Fegas} S. Nakatsuji {\it et al}., Phys. Rev. Lett. {\bf 99}, 157203 (2007).
\bibitem{Pauling} L. Pauling, {\it The Nature of the Chemical Bond} (Cornell University Press, Ithaca, NY, 1974).
\bibitem{CoIsing} M. Date, and M. Motokawa, Phys. Rev. Lett. {\bf 16}, 1111 (1966).
\bibitem{Takubo} K. Takubo, and T. Mizokawa, (unpublished).
\bibitem{Tsunetsugu} H. Tsunetsugu, and M. Arikawa, J. Phys. Soc. Jpn. {\bf 75}, 083701 (2006).
\bibitem{Penc} A. L\"auchli, F. Mila, and K. Penc, Phys. Rev. Lett. {\bf 97}, 087205 (2006).
\bibitem{Senthil} S. Bhattacharjee, V.B. Shenoy, and T. Senthil, Phys. Rev. B {\bf 74}, 092406 (2006).
\bibitem{Shen} P. Li, G.M. Zhang, and S.Q. Shen, Phys. Rev. B {\bf 75}, 104420 (2007).
\bibitem{KY} H. Kawamura, and A. Yamamoto, J. Phys. Soc. Jpn. {\bf 76}, 073704 (2007).
\bibitem{Fujimoto} S. Fujimoto, Phys. Rev. B {\bf 73}, 184401 (2006).
\bibitem{AKLT} I. Affleck, T. Kennedy, E.H. Lieb, and H. Tasaki, Phys. Rev. Lett. {\bf 59}, 799 (1987).
\bibitem{MSE} K. Yosida and S. Inagaki, J. Phys. Soc. Jpn. {\bf 50}, 3268 (1981).
\end{thebibliography}
\end{document}